\documentclass[mathleft
]{an}
\usepackage{graphicx}
\usepackage{times}
\begin{document}

\Pagespan{1}{}
\Yearpublication{}%
\Yearsubmission{}%
\Month{}%
\Volume{}%
\Issue{}%

\title{Imaging the cool stars in the interacting binaries AE Aqr, BV Cen
and V426 Oph}

\author{C.A. Watson\inst{1}\fnmsep\thanks{Corresponding author:
  \email{c.watson@sheffield.ac.uk}\newline}
\and  D. Steeghs\inst{2} \and V.S. Dhillon\inst{1} \and T. Shahbaz\inst{3}
}
\titlerunning{Imaging the cool stars in interacting binaries}
\authorrunning{C.A. Watson, D. Steeghs, V.S. Dhillon \& T. Shahbaz}
\institute{
Department of Physics and Astronomy, University of Sheffield, Sheffield, S3 7RH, UK
\and 
Harvard-Smithsonian Center for Astrophysics, 60 Garden Street, Cambridge, MA 02318, USA
\and 
Instituto de Astrof\'{i}sica de Canarias, 38200 La Laguna, Tenerife, Spain}

\received{}
\accepted{}
\publonline{later}


\abstract{It is well known that magnetic activity in late-type stars
increases with increasing rotation rate. Using inversion techniques
akin to medical imaging, the rotationally broadened profiles from such
stars can be used to reconstruct `Doppler images' of the distribution
of cool, dark starspots on their stellar surfaces. Interacting
binaries, however, contain some of the most rapidly rotating late-type
stars known and thus provide important tests of stellar dynamo
models. Furthermore, magnetic activity is thought to play a key role
in their evolution, behaviour and accretion dynamics. Despite this, we
know comparatively little about the magnetic activity and its
influence on such binaries. In this review we summarise the concepts
behind indirect imaging of these systems, and present movies of the
starspot distributions on the cool stars in some interacting
binaries. We conclude with a look at the future opportunities that
such studies may provide.}

\maketitle

\section{Introduction}

Starspots, the most easily observed manifestation of magnetic
activity, are a ubiquitous feature of rotating late-type stars with
outer convective zones. Indeed, naked-eye observations of spots on our own Sun
first took place 1000's of years ago, and subsequent telescopic
observations have been undertaken almost continuously for over 3
centuries. Despite this, our understanding of how
magnetic fields are produced and evolve on the Sun and other stars,
and how fundamental parameters affect magnetic field generation, is
still critically lacking.  Since Vogt \& Penrod (1983) demonstrated
that starspots could be seen as bumps traversing the line profiles of
HR 1099, several indirect imaging techniques have been used to map the
magnetic topographies on mainly (but not exclusively) isolated rapid
rotators. Since it is well known that magnetic activity increases with
increasing stellar rotation rate, much of this work has focused on
determining how rotation impacts starspot emergence.

Interacting binaries such as Cataclysmic Variables (CVs) contain some
of the most rapidly rotating late-type stars known. CVs typically
consist of a late-type main-sequence star in orbit around a more
massive white dwarf. The close proximity of both components causes the
late-type (or donor) star to fill its Roche-lobe and
transfer mass to a white dwarf. With orbital periods as
short as an hour, and tides forcing synchronous rotation of the
donor, images of these systems provide excellent tests of
stellar dynamo theories at ultra-fast rotation rates. In addition,
such systems also allow the impact of tides on the stellar dynamo to
be assessed.

Furthermore, many interacting binaries show variations in orbital
periods, brightnesses, and outburst durations that have been
attributed to starspots and activity cycles. Despite their obvious
importance, both to understanding stellar dynamos and  understanding
the observed behaviour of interacting binaries, little progress in
this field has been made until recently. In this review, we will
describe the astro-tomographic techniques required to image starspots
on the cool stars in interacting binaries, before summarising some of
the recent highlights in this field. Finally, we will conclude with
some future opportunities.

\section{The technique of Roche Tomography}

Unfortunately, interacting binaries are too small and too distant to
be resolved directly, even in the worlds largest telescopes. Rutten \&
Dhillon (1994), however, described a technique, called Roche
tomography, which uses phase-resolved spectra to reconstruct the
line-intensity distribution on the donor stars.  While analogous to
Doppler imaging methods applied to single stars, Roche tomography
differs in two important ways. First, the donor stars are tidally
distorted into a Roche-lobe and are in synchronous rotation around the
centre-of-mass of the binary. Second, the systems often show rapid
variability due to the accretion. This means
that one usually requires simultaneous photometry with which to
slit-loss correct the spectra before continuum subtraction, whereas
with Doppler imaging the spectra need not be slit-loss corrected and
the continuum is normalised.

Otherwise, the concept of Roche tomography is relatively straight forward.
Given the binary parameters, the donor's surface can be
defined to lie on the critical surface of equal gravitational
potential given by the Roche approximation at the point where
Roche-lobe overflow occurs. The surface can then be modelled as a
series of tiles lying on this critical
surface. Each tile is then assigned a copy of the
local specific intensity profile (which may either be Gaussian or
taken from a spectral-type standard) and convolved with the
instrumental resolution. One can then assign an intensity to each
tile which scales its contribution to the total profile.  Each
tile's contribution is then integrated over the visible surface of
the star taking into account limb-darkening, fore-shortening and the
radial-velocity of the elements.

Roche tomography essentially carries out the reverse of the process
just described. The contribution of each element is iteratively varied
until a map is obtained which predicts data consistent with the
observed data.  The goodness of fit is measured using the reduced
$\chi^2$ statistic. Unfortunately, the reduced $\chi^2$ constraint
alone is not sufficient to select a unique map as there are many maps
that can fit the data equally well, and so we also adopt an additional
{\em regularisation} statistic. Following Horne (1985), we adopt the
map of {\em maximum entropy}, which can loosely be thought of as the
smoothest map, or map containing the least information, with respect
to a {\em default map} which may contain some a priori information
about the surface intensity distribution across the donor
star. Generally, we assume a uniform default map whose pixels are set
to the average value in the reconstructed map, and the default map is
updated after each iteration. An efficient algorithm for maximising
entropy subject to a $\chi^2$ constraint has been implemented by
Skilling \& Bryan (1984) in the {\sc fortran} package {\sc memsys}.

\section{Roche Tomography highlights}

The earliest work using Roche tomography concentrated on
reconstructing surface maps of CV donors using either single
absorption or emission lines obtained using intermediate resolution
spectrographs. The maps clearly showed the varying impact of
irradiation from the compact object and accretion regions on these
stars. While interesting in their own right, these results have been
covered in several papers and reviews already and, instead of
repeating the findings here, we refer the reader to the work by
Watson et al. (2003). Instead, we shall focus on some of the more
recent results showing the starspot distribution on a number of CV
donors.

\subsection{AE Aqr -- the first image of starspots on a CV donor star}

Early, single line CV studies failed to reveal the presence of cool
starspots on CV donors since they neither had the resolution nor the
signal-to-noise to detect the small, faint signatures of starspots.
(One should consider that most CVs have $m_v >$11, generally about
2--4 mags fainter than most stars that have been Doppler imaged). The
first of these problems can be solved by employing a high resolution
echelle spectrograph. Simply moving to a larger aperture telescope,
however, does not solve the signal-to-noise problem.

Fortunately, the wide wavelength coverage of an echelle spectrograph
allows 1000's of stellar lines to be observed in one spectrum. One can
then apply a technique known as {\em Least Squares Deconvolution} (LSD
-- see Donati et al. 1997) which effectively combines all the lines to
form one `average' stellar absorption line of greatly increased
signal-to-noise.  Typically, this technique provides a multi-plex gain
in signal-to-noise of $\sim$30. To put this in perspective, using LSD
on an 8-m class telescope is the equivalent of attempting a single
line study on a 40-m class telescope!

Armed with this technique, we obtained two nights echelle spectroscopy
with the 4.2-m William Herschel Telescope (WHT) on La Palma on AE Aqr
as part of a pilot study. This target was chosen as it is the
brightest CV in the Northern sky, and the donor star contributes a
large fraction ($\sim70\%$) to the total system light. AE Aqr also has
a relatively long-orbital period of 9.87-hrs which meant that
reasonably long exposures could be used before the donor's orbital
motion started to smear out the line profiles. Application of LSD to
this dataset resulted in a multi-plex gain in signal-to-noise of
$\sim$26 over single-line studies.

\begin{figure}
\includegraphics[width=82mm]{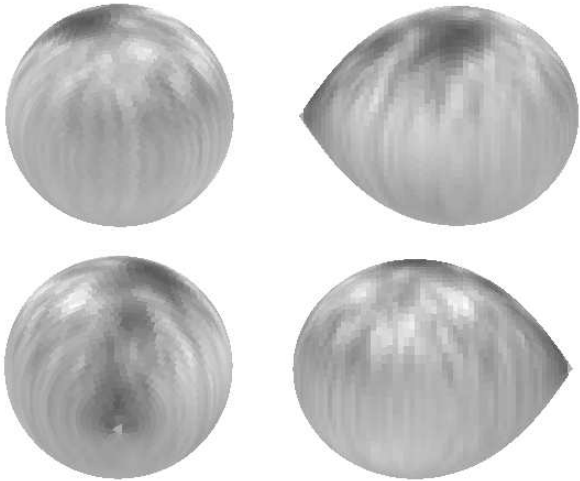}
\caption{The Roche tomogram of AE Aqr. The panels (clockwise from
top-left) show the cool donor star at orbital phases $\phi$=0
(inferior conjunction of the donor), $\phi$=0.25, $\phi$ = 0.75, and
finally $\phi$=0.5 (superior conjunction of the donor star). Dark
greyscales indicate regions that are either covered in starspots or
are irradiated. The system is plotted as the observer would view AE
Aqr at an orbital inclination of $i$ = 66$^{\circ}$.}
\label{fig:aeaqr}
\end{figure}

The Roche tomogram of AE Aqr is published in Watson et al. (2006), and
can be viewed as a movie on AN's webpages. Representative frames are
displayed in Figure~\ref{fig:aeaqr} and show several dark starspot
features.  The most prominent of these is a large, high latitude spot
most clearly visible at orbital phase $\phi$=0.25. This spot is
centred on a latitude of $\sim$65$^{\circ}$, in stark contrast to our
Sun where few spots are seen above latitudes greater than
$\sim$35$^\circ$. Such a high latitude feature on AE Aqr is, however,
consistent with Doppler images of rapid rotators which often show the
presence of polar spots.

Another feature of note is the dark region surrounding the $L_1$
point, which we have interpreted as irradiation of the inner face by
the white dwarf and/or accretion regions. Such irradiation zones in
similar locations in CV donors have been mapped previously (e.g. Davey
\& Smith 1996; Watson et al. 2003). There does, however, appear to be
a chain of spots extending from the polar regions down to the
irradiated $L_1$ point, making longitudes of the star facing the white
dwarf companion appreciably more spotted than elsewhere. Indeed, in
the Doppler image of the pre-CV V471 Tau, Hussain et al. (2006) also
found that the side of the star facing the white dwarf was heavily
spotted. This may well be due to tidal forces, which have been
predicted to cause spots to emerge at preferred longitudes
(e.g. Holzwarth \& Sch{\"{u}}ssler 2003).

In total, Watson et al. (2006) estimated that 18 per cent of AE Aqr's
Northern hemisphere was covered with spots, providing the first
conclusive evidence that these systems were magnetically active.  This
put the canonical theory of CV evolution (which required strong
magnetic fields on the donors in order to drain angular momentum from
the binary and drive the systems to shorter orbital periods) on a more
secure observational footing.

\subsection{BV Cen -- starspots and slingshot prominences}

\begin{figure}
\includegraphics[width=79.7mm]{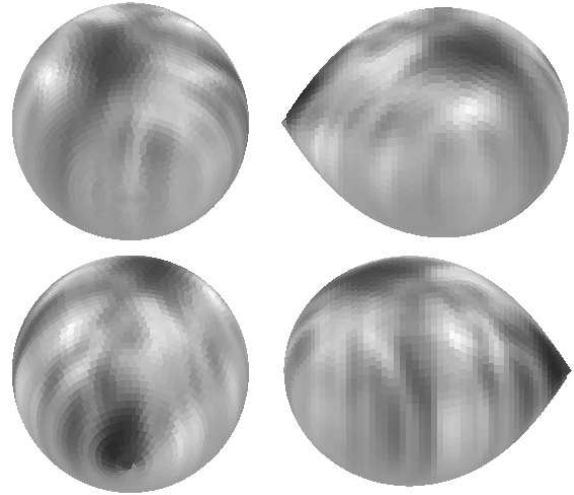}
\caption{The same as Fig.~\protect\ref{fig:aeaqr}, but for BV Cen and
assuming an orbital inclination of $i$ = 53$^{\circ}$.  While BV Cen
clearly has a larger spot coverage than AE Aqr, the similarities in
the surface distribution of spots between BV Cen and AE Aqr is
striking.}
\label{fig:bvcen}
\end{figure}

Based on the success of the AE Aqr pilot study we began a campaign on
the 6.5-m Magellan Clay telescope situated in Chile to map other CV
donors. The larger aperture, wider spectral coverage and higher
throughput of the Magellan setup compared to the WHT allowed systems 2
or 3 magnitudes fainter then AE Aqr to be imaged. The second CV we
observed was the $\sim$14.7 hour period system BV Cen.  A movie of BV
Cen, reconstructed after applying LSD, is available on the AN website
and some snapshots are shown in Figure~\ref{fig:bvcen}.

The maps of BV Cen and AE Aqr (Figure~\ref{fig:aeaqr}) make for
interesting comparison. Again, BV Cen shows a high latitude spot at
$\sim$65$^{\circ}$. Interestingly, the high latitude spot in both
systems are displaced in the same direction towards the trailing
hemisphere of the star (the side facing away from the orbital motion
of the star). While we believe that the prominent dark region near the
$L_1$ point is due to irradiation, we also see a chain of spots
extending from the polar regions down to the $L_1$ point similar to
that seen on AE Aqr. This provides some evidence that a mechanism is
at work, perhaps tidal and/or Coriolis forces, that is forcing flux
tubes to arise at these locations.

The fact that regions near the $L_1$ point seem heavily spotted,
combined with the fact that low latitude spots are seen near  the
$L_1$ point, has interesting implications for the accretion dynamics
of these systems. Many accreting binaries show sudden dips in their
lightcurves, and it has been suggested that this is caused by
starspots moving across the $L_1$ point and quenching the mass
transfer, leading to a dimming in the system's brightness.  In their
modelling of the mass transfer history of AM Her, Hessman,
G{\"{a}}nsicke \& Mattei (2000) concluded that if starspots were to
cause such dips then regions near the $L_1$ point would most likely be
unusually heavily spotted, with a spot-filling factor around
50\%. Certainly, both AE Aqr and BV Cen appear to be more densely
spotted on the hemisphere facing the white-dwarf. Indeed, we have
calculated that, for BV Cen, the spot-filling factor reaches 40\% at
longitudes near the $L_1$ point, which seems to support the
conclusions of Hessman et al. (2000).

In addition to the presence of starspots in BV Cen, which we
calculated covered some 25\% of the Northern hemisphere, we also
detected a transient narrow emission feature at zero velocity. Such
emission has been seen previously in other CVs (see e.g. Steeghs et
al. 1996), also situated at zero velocity, and have been interpreted
as `slingshot prominences' from the donor star sitting near the
centre-of-mass of the binary. The preference for prominences to be
seen at this location may be a combination of two things. First,
prominences erupting from near the $L_1$ point will be more subject to
illumination from the accretion regions, making them more visible to
us. Second, it appears from the Roche tomograms of BV Cen and AE Aqr
that spots are more likely near the $L_1$ point, which may increase
the probability of a prominence erupting in these active regions.
Furthermore, with such a strong concentration of activity near
the $L_1$ point, magnetic fragmentation of the accretions stream
may be responsible for the `blobby' accretion
seen in some CVs (e.g. Meintjes 2004; Meintjes \& Jurua 2006).

\subsection{V426 Oph -- caught during a prolonged low-state}

\begin{figure}
\includegraphics[width=79mm]{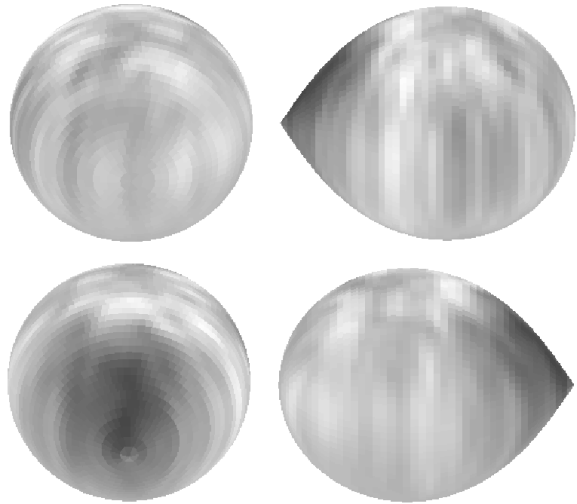}
\caption{The same as Figures~\protect\ref{fig:aeaqr}
and~\protect\ref{fig:bvcen} but for V426 Oph, assuming an orbital
inclination of 63$^{\circ}$.}
\label{fig:v426oph}
\end{figure}

Finally, we present a preliminary Roche tomogram of V426 Oph, again as
a movie with some representative frames shown in
Figure~\ref{fig:v426oph}. With an orbital period of $\sim$6.85 hrs
this is, to our knowledge, the fastest rotating star on which
starspots have been imaged to date, the previous record being 7.44 hrs
(RXJ1508.6 -- Donati et al. 2000).

Unlike the Roche tomograms of AE Aqr and BV Cen, V426 Oph does not
show such a prominent high latitude spot, though there is still a hint
of a weak polar feature, again displaced slightly towards the trailing
hemisphere.  V426 Oph also shows spots appearing at lower
latitudes. With a rotation rate almost 90 times solar, this is at odds
with the models of Sch{\"{u}}ssler et al. (1996) which predict that
magnetic flux tubes should only break the surface at high latitudes
due to Coriolis forces and magnetic buoyancy.

Although the reason for the lack of a high latitude spot is
unknown, one possibility is that we caught it at an unusual stage in
its activity cycle. Interestingly, data from the AAVSO
indicates that V426 Oph was undergoing some unusual behaviour at this
time -- being both dimmer during 2005 (when this data was taken),
and exhibiting far fewer outbursts compared to other years. Was V426
Oph in an unusual state of activity, and is this responsible for its
unusual behaviour during 2005? Further observations are
required if any such link is to be made, or if we are to determine
whether V426 Oph simply lacks a polar spot at all times.

Other than that, the inner face of V426 Oph appears to be quite
uniformly irradiated (giving the star a darker appearance in the Roche
tomograms at $\Phi$=0.5).  Within the irradiated zone, however,
individual starspots are still visible near the $L_1$ point, and V426
Oph also appears to exhibit a less distinctive chain of spots extending
from high latitudes down to the $L_1$ point. Given the preliminary
nature of the analysis presented here, we are hesitant to draw any
more conclusions at this time.

\section{Future opportunities}

While we are now beginning to reveal in some detail the magnetic
nature of CVs, a sustained campaign is required to investigate whether
these systems show solar type activity cycles, and to what extent
such a cycle impacts on the observed behaviour of the binary. We are
currently undertaking a monitoring campaign to determine whether
AE Aqr displays any solar-like activity cycle.

Another interesting question is whether the donor stars in interacting
binaries display an appreciable level of differential
rotation. Scharlemann (1982) suggests that tidal
forces should weaken, but not suppress, differential rotation. The
fact that we see very similar spot distributions on AE Aqr and BV Cen
suggests that the spot distribution may essentially be fixed. In turn,
this suggests that differential rotation on these systems may be weak,
perhaps even suppressed. Given that differential rotation is a key
ingredient in stellar dynamo models, amplifying and transforming
initial poloidal field into toroidal field, any suppression of
differential rotation may provide a challenge to dynamo theory. Thus
measurements of the differential rotation rate, via cross-correlation
of latitude strips from images taken 1 or 2 weeks apart, is
essential. Such studies on binaries have
a unique advantage over those conducted on single stars in
that the {\em sense} of the differential rotation can be deduced since
the actual mean rotation period is known from the orbital period of
the binary, allowing the corotation latitude of the donor
star to be determined.

Finally, it would be interesting to prove that starspots can affect
the accretion dynamics of interacting binaries by suppressing mass
transfer. An observation of a starspot crossing the $L_1$ point,
combined with a dip in the luminosity of the system, would be
compelling evidence for such a scenario. It is quite comforting to see
that, in the case of V426 Oph, we can still identify starspots even if
they lie within irradiated regions. In summary, such work promises to
reveal much about both the behaviour of binaries, and of the stellar
dynamo under the most extreme conditions.

\end{document}